\documentclass[12pt]{elsart}

\usepackage{graphicx}
\usepackage{longtable}

\usepackage{amsmath, amssymb}
\usepackage{color}

\newcommand{\nl}{\nonumber\\ }
\newcommand{\pd}{\partial}
\newcommand{\TGint}{\int\mathrm d\Gamma}

\newcommand{\scid}{^{sc}}

\newcommand{\avgint}[1]{\left\langle{#1}\right\rangle}

\def\be{\begin{eqnarray}}
\def\ee{\end{eqnarray}}

\def\lsim{\stackrel{\scriptstyle <}{\phantom{}_{\sim}}}
\def\gsim{\stackrel{\scriptstyle >}{\phantom{}_{\sim}}}
\def\rmd{{\rm d}}

\begin{document}
\begin{frontmatter}

\title{Remarks concerning bulk viscosity of hadron matter in relaxation time ansatz}

\author[JINR,Mold]{A.S.~Khvorostukhin},
\author[JINR]{V.D.~Toneev},
\author[MEPHI]{D.N.~Voskresensky}
\address[JINR]{Joint Institute for Nuclear Research,
 141980 Dubna, Russia}
\address[Mold]{ Institute of Applied Physics, Moldova Academy of Science,
MD-2028 Kishineu, Moldova}
\address[MEPHI]{National Research Nuclear University "MEPhI",
Kashirskoe sh. 31, Moscow 115409, Russia}




\begin{abstract}
The bulk  viscosity is calculated for  hadron matter produced in
heavy-ion collisions, being described  in the relaxation time
approximation within the relativistic mean-field-based model with
scaled hadron masses and couplings. We show how different
approximations used in the literature affect the result. Numerical
evaluations of the bulk viscosity with three considered
models
 deviate not much from each other confirming earlier results.
\end{abstract}
\end{frontmatter}

\maketitle

\section{Introduction} Recently, interest in the transport
coefficient issue for hadronic and quark matter  has essentially
increased due to clarifying the role of viscosity in extraction of
flow parameters from heavy-ion
collisions, see review-article~\cite{K08}.  Viscosity
coefficients in a weakly coupled scalar field theory at an
arbitrary temperature can be evaluated directly from the first
principles using expansion of the Kubo formulas in terms of ladder
diagrams in the imaginary time formalism \cite{Je94}.
Unfortunately, similar analysis for more general cases is
unavailable. Therefore, in order to evaluate transport coefficients
in multi-component systems with a strong coupling between species,
one often uses a kinetic approach. Thus, one can exploit  the
relaxation time approximation to the Boltzmann-like quasiparticle
kinetic equations. The shear and bulk viscosities of the hadron
and the quark-gluon plasma phases of strongly interacting matter
at finite temperature and baryon density were evaluated, see
Refs. \cite{SR08,SHMCbulk,CK10}. In~\cite{GBulk,GBulk1,BKR10}, a
similar analysis was  performed
for purely gluon matter. Besides the use of the relaxation time
approximation,  one needs to do some extra assumptions in order
to proceed further. In~\cite{GBulk1}, we studied how different ansatze
used in the literature affect the result for shear and bulk
viscosities for  gluon matter. We found that the result for the
shear viscosity was robust with respect to different ansatz reductions, whereas
the value of the bulk viscosity significantly depends on them.
 In this note
we study how different approximations used in the literature
affect the bulk viscosity of the hadron matter.

\section{ Model equations}  As in \cite{SHMCbulk}, we describe the
hadron phase in terms of the quasiparticle relativistic mean-field
(RMF)-based model with the scaling hadron mass-couplings (SHMC)
successfully applied earlier to the description of heavy-ion
collision reactions ~\cite{SHMCmodel}. The model used is an extension of
the model~\cite{KV04} applied there for the cold dense hadron
matter.
Bearing in mind its application to heavy-ion collisions, we deal
with a RMF-based  model of iso-symmetric non-equilibrium hadron
matter with $\sigma$- and $\omega$-meson mean fields and in
contrast with  ordinary RMF models we assume that  not only baryon
but also other hadron masses might depend on the $\sigma$-meson
mean field. 
Also excitations emerging from the $\sigma$ and $\omega_0$ mean
fields are incorporated.
 We study the system with zero net strangeness and  use the
 same hadron set, as in \cite{SHMCmodel,SHMCbulk}.
Considering small deviations from local equilibrium we keep only
first-order gradient terms. Further details can be found in the
mentioned works \cite{SHMCmodel,SHMCbulk}.

 We start with the Lagrangian of SHMC model \cite{SHMCmodel}
 from where  expressions are reproduced for the  baryon/strangeness
 4-current and the energy-momentum tensor densities
 \cite{SHMCbulk}:
\begin{eqnarray}
\label{current} J^\mu_{B,S}&=&\sum_a t_a^{B,S}\TGint_a
\frac{p^\mu_a}{E_a}\, f_a(x,\vec{p})~,\\
 T^{\mu\nu}&=&\sum_a\TGint_a \frac{(p_a^\mu +X_a^\mu) p^{\nu}_a} {E_a}\, f_a(x,\vec{p})
 +T^{\mu\nu}_{\rm MF}
~.\label{Tmunu}
 \end{eqnarray}
Here $ \rmd\Gamma_a=d_a\frac{\rmd^3 p}{(2\pi)^3}~,\quad p_a^\mu =
\left(E_a ,\vec{p}\right)~,$
\begin{eqnarray}\label{X0}
   E_a =\sqrt{\vec{p}^{\,2}+m_a^{*\,2}(\sigma)}~,\quad
   X_a^\mu = X^0_a(\sigma,\omega_0)\,\delta^{\mu0}~,
\end{eqnarray}
 $t_a^B,\ t_a^S$ are
the  baryon and strange charges of the $a$-hadron (antiparticles
are included), $d_a$ is the degeneracy factor, $m_a^{*}(\sigma)$ is
the effective mass, $f_a$ is the quasiparticle distribution function,
$\sigma$ and $\omega_0$ are mean scalar and vector meson fields,
$\omega^\mu =(\omega_0,\vec{0})$, $X_a^0\sim\omega_0$,
\begin{eqnarray}\label{TmunuMF}
T^{\mu\nu}_{\rm MF}&=&g^{\mu\nu}\left[
 U(\sigma)-V(\sigma,\omega)\right],\\
 U(\sigma)&=&\frac{m_{\sigma}^{*\,2}(\sigma)\sigma^2}{2}+U_{\rm
NL}(\sigma),\nonumber\\
V(\sigma,\omega)&=&\frac{m_{\omega}^{*\,2}(\sigma)\omega_0^2}{2},\nonumber
  \end{eqnarray}
where  $U_{\rm NL}(\sigma)$ is the non-linear potential of the $\sigma$ field.
In general, since we derive (\ref{Tmunu}) from the
Lagrangian, in our mean-field approach  the energy-momentum tensor
can be symmetrized following the standard rule, e.g. see
\cite{Weinberg}. In this note we  study viscosities.
 For their calculation  we need spatial components of the tensor $T^{ik}$:
\begin{eqnarray}\label{sym}T^{ik}=\sum_a\TGint_a \frac{p_a^i  p^{k}_a} {E_a}\, f_a(x,\vec{p})
 +T^{ik}_{\rm MF}\,.
\end{eqnarray}
 Latin indices $i,k=1,2,3$. This term is symmetric. Here we used that in the mean-field
approximation, which we exploit, only $X_0\neq 0$ in local
equilibrium, see (\ref{X0}). This causes contributions $\propto
\partial {u}_j/\partial x_k$,
when we further calculate viscosities, and $u_j $ is $j$-spatial
component of the hydrodynamic  velocity.  The spatial components
$X_j $ yield only higher  order  terms in $\partial {u}_j/\partial
x_k$ in the local rest frame. Therefore, their contribution can be
omitted.

   Conservation laws of
the baryon/strangeness 4-current and of the energy-momentum tensor
densities are read as
\begin{eqnarray}\label{consT}
\pd_\mu J_{B,S}^\mu&=&0~,\quad\partial_\mu T^{\nu\mu}=0~.
\end{eqnarray}

We assume that the quasiparticle distribution functions $f_a$ obey
a set of kinetic equations
\begin{eqnarray}\label{kineq}
   E_a^{-1} p_a^\mu\pd_\mu f_a -\nabla E_a \nabla_{{p}} f_a &=& \mbox{St} f_a~,
\end{eqnarray}
where $\mbox{St} f_a$ is the collision term for the given species
satisfying the conditions
\begin{eqnarray}\label{consSt}
\sum_at_a^{B,S}\TGint_a{\rm St}f_a=0,\quad
\sum_a \TGint_a\varepsilon_a{\rm St}f_a=0,
\end{eqnarray}
with the quasiparticle energy $\varepsilon_a=E_a+X_a^0$.
 Further in the Boltzmann equations
(\ref{kineq}) we omit  the term  $\propto\nabla E_a$, which does
not  contribute to viscosities.

 The collision term is zero for the  local
equilibrium distribution,  ${\rm St}f_a^{\rm l.eq.}=0$,
 where
\begin{eqnarray} \nonumber
&& f_a^{\rm l.eq.} (E_a^{\rm l.eq.},\vec{p}\, ,x^\mu ) \\
&=&
 \left\{e^{[E_a^{\rm l.eq.}-\vec{p}\,\cdot \vec{u}(x^\mu)-\mu^*_a(x^\mu)]/T(x^\mu)}\pm 1
 \right\}^{-1},
 \label{leqdf}
\end{eqnarray}
with $+$ for fermions, $-$
for bosons,
\begin{eqnarray*}
E^{\rm l.eq.}_a&=&E_a(\sigma^{\rm l.eq.})
~,\\ \mu^*_a(x^\mu)&=&t_a^B
\mu_B(x^\mu)+t_a^S\mu_S(x^\mu)-X^0_a(\sigma^{\rm
l.eq.},\omega_0^{\rm l.eq.})~,
\end{eqnarray*}
$\sigma^{\rm l.eq.}=\sigma^{\rm l.eq.}(T,\mu_B,\mu_S)$,
$\omega_0^{\rm l.eq.}=\omega_0^{\rm l.eq.}(T,\mu_B,\mu_S)$,
$\mu_B,\mu_S$ are the baryon and strangeness chemical potentials,
and the four-velocity of the frame is $u^\mu \simeq [1,
\vec{u}(x^\mu)]$ for $|\vec{u}|\ll 1$.

 Applying the second Eq.~(\ref{consT}) for $\nu =0$ with  the condition (\ref{consSt}),  and
using~(\ref{Tmunu}) we derive the self-consistency conditions:
\begin{eqnarray}
\label{selfcons1}
  \frac{\pd V}{\pd\omega_0}&=&\sum_a\TGint_a f_a\,\frac{\pd\varepsilon_a}{\pd\omega_0}~,\\
  \label{selfcons2}
  \frac{\rmd  U}{\rmd \sigma}-\frac{\pd V}{\pd \sigma}&=&-\sum_a\TGint_a
  f_a\,\frac{\pd\varepsilon_a}{\pd\sigma}~.\nonumber
\end{eqnarray}
 For the  equilibrium system  these equations  coincide with
the conditions of maximum pressure $ \pd
P/\pd\omega_0=0,\quad\pd P/\pd\sigma=0$, where pressure
$P=\frac{1}{3}T_{ii}^{\rm l.eq.}$. 

 In the general case, it is impossible to solve the Boltzmann kinetic
 equations for  the strongly interacting
multi-hadron system appearing in the course of heavy-ion
collisions. However the collision term is greatly simplified in
the so-called relaxation time approximation or more precisely, in
the relaxation time  ansatz.
{\em{ Near the local
 equilibrium state we will use the expansion }}
 \begin{eqnarray}\label{trel}
 {\rm{St}}
f_a =-\delta f_a/\tau_a  , \quad \delta f_a =f_a -f_a^{\rm
l.eq.}(E_a^{\rm l.eq.}),
\end{eqnarray}
where $\tau_a$ are in general the energy-dependent quantities,
i.e., $\tau_a = \tau_a (E_a^{\rm l.eq.})$. These values can be
evaluated from the cross sections of particle-particle
interactions.

Since within the relaxation time
approximation the expression for the shear viscosity, $\eta$, is easily
recovered  \cite{SHMCbulk}  and one  needs no extra assumptions for that,
we study below the bulk viscosity, $\zeta$, only.

\section{Bulk viscosity}
  The bulk viscosity is defined as the coefficient
entering into the variation of  $T^{ii}$ in the local rest frame:
\begin{eqnarray}
 \label{vis}
 {\delta T^{ii}}/{3}&=&-\zeta \ \nabla\cdot\vec{u}~,
\end{eqnarray}
and  variations of the baryon/strange charge and the energy
density should satisfy the so-called Landau-Lifshitz 
matching conditions  $u_\mu \delta J^{\mu}_{B,S}=0$ and $u_{\mu}
\delta T^{\mu \nu}u_{\nu} =0$, which in the local rest frame
reduce to 
\begin{eqnarray}
 \label{LLgen}\delta J^0_{B,S}&=&\sum_a  t_a^{B,S}\TGint_a\,\delta f_a=0~,\\
\delta T^{00}&=&\sum_a\TGint_a\varepsilon_a\,\delta f_a=0
~.\label{LLgen1}
\end{eqnarray}
 These conditions are necessary to make the system
thermodynamically stable in the first-order theory \cite{Hirano}.

 From the Boltzmann
equations within the relaxation time approximation we find
\begin{eqnarray}\label{delf} \delta f_a
[{{\nabla}}\cdot\vec{u}]
&=&\left[\tau_a
Q_a(\vec{p}^{\,2})\frac{f_a (1\mp f_a)}{T}\right]^{\rm
l.eq.}{{\nabla}}\cdot\vec{u}~,
\end{eqnarray}
 where
\begin{eqnarray}
\label{defQa} -Q_a(\vec{p}^{\,2})
&=&\frac{\vec{p}^{\,2}}{3E_a}+\left(\frac{\pd
P}{\pd\epsilon}\right)_{n_B,n_S}\nl
&&\times\left(T\frac{\pd\varepsilon_a}{\pd T}+\mu_B\frac{\pd
\varepsilon_a}{\pd\mu_B}+\mu_S\frac{\pd \varepsilon_a}{\pd\mu_S}-
\varepsilon_a\right)\nl &&+\left(\frac{\pd P}{\pd
n_B}\right)_{\epsilon,n_S}\left(\frac{\pd
\varepsilon_a}{\pd\mu_B}-t_a^B\right)\nl &&+\left(\frac{\pd P}{\pd
n_S}\right)_{\epsilon,n_B}\left(\frac{\pd
\varepsilon_a}{\pd\mu_S}-t_a^S\right)
\end{eqnarray}
and $\epsilon=T_{00}^{\rm l.eq.}$, cf. \cite{SR08,SHMCbulk}. We
retained only terms with $\nabla \vec{u}$ since now we are
interested in the calculation of the  bulk viscosity.

Note that with $\delta f_a$ obeying Eq. (\ref{delf}) we are
able to explicitly show that initially asymmetric contribution to
the energy-momentum tensor $T^{0i}$ is zero. Indeed,
\begin{align} \Delta T^{0i}&=\sum_a X_a^0\TGint \frac{
p^i}{E_a}\,\delta f_a =0\,
\end{align}
due to angular integrations. Thus our initially asymmetric
expression for the energy-momentum tensor (\ref{Tmunu}) does not
cause any problems in calculation of viscosity.

 In the relaxation time approximation,
 using (\ref{delf}) we present the Landau-Lifshitz
conditions (\ref{LLgen}), (\ref{LLgen1}) as
\begin{eqnarray}
 \label{LLaI2} \sum_a t_a^{B,S}\avgint{\tau_a Q_a(\vec{p}^{\,2})}&=&0~,\\
\label{LLaI1} \sum_a\avgint{\tau_a\varepsilon_a Q
_a(\vec{p}^{\,2})}&=&0 ~.
\end{eqnarray}
Here  we introduced  notation
\begin{eqnarray}
\left\langle \Phi_a(\vec{p}) \right\rangle =\TGint_a\,
\left[\Phi_a(\vec{p})f_a(1\pm f_a)\right]^{\rm l.eq.}~.
\end{eqnarray}

Similarly Eqs. (\ref{consSt}) are reduced to
\begin{eqnarray}
 \label{LLaI2a} \sum_a t_a^{B,S}\avgint{Q_a} =0~,\quad
\sum_a\avgint{\varepsilon_a Q _a} =0 ~.
\end{eqnarray}
Using standard thermodynamic relations and self-consistency
relations (\ref{selfcons1}),
one may show that Eqs. (\ref{LLaI2a}) are indeed fulfilled.

 To continue calculation of the bulk viscosity
additional approximations are needed. Below we introduce three
possible ansatze  and  compare results of calculations.

\subsection{ Model I (Refs. \cite{SR08,SHMCbulk}).} Following the
line sketched in Ref. \cite{SR08}, performing variations  in Ref.
\cite{SHMCbulk} {\em{ we did not vary quantities which may depend
on the distribution function only implicitly}}, such as $E_a$.
This approximation  is well satisfied for non-relativistic
systems, see \cite{GBulk1}.  Although  the validity of this
approximation becomes questionable in the application to
relativistic systems, its use allows one to essentially simplify
calculations for the bulk viscosity, which is important in the
case of  a complicated system of many strongly interacting
particle species. Therefore, in~\cite{SHMCbulk} we
used it as an additional
 assumption.  Then the expression for the  value $\delta T^{ii}$
looks very simple
\begin{eqnarray}
 \label{deltaPaIII}
  \delta T^{ii}[\delta f_a]&=&\sum_a\TGint_a \frac{\vec{p}^{\,2}}{E_a}\,\delta
  f_a~,
\end{eqnarray}
since $E_a$ values are  not varied. Using (\ref{deltaPaIII})  we arrived
at the following expression for the bulk viscosity:
\begin{eqnarray}\label{viskold}
  \zeta &=& -\frac1T\sum_a\avgint{\tau_a
  Q_a(\vec{p}^{\,2})\frac{\vec{p}^{\,2}}{3E_a}}~.
\end{eqnarray}

Using the Landau-Lifshitz condition (\ref{LLaI1}) we present
Eq. (\ref{viskold}) in the form ~\cite{SHMCbulk}
\begin{eqnarray}\label{bulkSHMC}
  \zeta _{\,\text{Ref.\cite{SHMCbulk}}}&=& \frac{1}{3T}\sum_a\avgint{\tau_a
  Q_a(\vec{p}^{\,2})\left(\frac{m_a^{*\,2}}{E_a}+X_a^0\right)}~.
\end{eqnarray}
 Here, as in~\cite{SR08}, {\em we have just assumed the validity of the
Landau-Lifshitz conditions (\ref{LLaI2}) and
(\ref{LLaI1}).} However one should note that these  relations
might not be fulfilled, until  some additional conditions were not
imposed, see below. Since we did not impose these extra
conditions, the use of the Landau-Lifshitz matching conditions can
be considered  just as an additional not yet justified assumption.
Nevertheless, as it follows from
(\ref{LLaI2a}), at least in simplest cases, like for a
one-component system with $\tau (E) =\,$const and, more generally,
for $\tau_a=\tau=\,$const, the Landau-Lifshitz conditions are
indeed fulfilled.

 Concluding,  we stress that all quantities in Eqs.
(\ref{viskold}) and (\ref{bulkSHMC}) including $E_a$ are taken at
local equilibrium.

\subsection{ Model II}
 We return to the relaxation time
ansatz. But now we avoid additional two assumptions used in
\cite{SR08,SHMCbulk}. To derive the expression for the bulk
viscosity, we follow the procedure sketched  in 
[7, Sec. III.A] for
gluons.

First, from (\ref{selfcons1}),
(\ref{selfcons2}) we find
variations of
 mean fields
\begin{eqnarray}\label{Lmatr}
\begin{pmatrix}
  \delta \sigma\\
   \delta\omega_0\\
\end{pmatrix}
=
\begin{pmatrix}
  \Lambda_\sigma & \Lambda_{\sigma\omega} \\
  \Lambda_{\sigma\omega}& \Lambda_\omega \\
\end{pmatrix}
\begin{pmatrix}
   \sum_a\TGint_a \frac{\pd\varepsilon_a}{\pd\sigma}\, \delta f_a\\
   \sum_a\frac{\pd X^0_a}{\pd\omega_0}\TGint_a \, \delta
  f_a\\
\end{pmatrix}~,
\end{eqnarray}
where
\begin{eqnarray}
  \Lambda_\sigma\Lambda &=&\sum_a\frac{\pd^2 X_a^0}{\pd\omega_0^2}\,n_a-\frac{\pd^2 V}{\pd\omega_0^2}~,\\
    \Lambda_{\sigma\omega}\Lambda &=&\frac{\pd^2 V}{\pd\omega_0\pd \sigma}-\sum_a\frac{\pd^2 X_a^0}{\pd\omega_0\pd
    \sigma}\,n_a~,\nonumber
\end{eqnarray}
\begin{eqnarray}
    \Lambda_\omega\Lambda &=&\frac{\rmd^2  U}{\rmd\sigma^2}-\frac{\pd^2  V}{\pd \sigma^2}
   +\sum_a\left(\frac{\rmd m_a^*}{\rmd\sigma}\right)^2\TGint_a\frac{\vec{p}^{\,2}}{E_a^3}\,f_a^{\rm l.eq.}
   \nl&&+\sum_a\frac{\rmd^2m_a^*}{\rmd\sigma^2}\,\rho\scid_a+\sum_a\frac{\pd^2
   X^0_a}{\pd\sigma^2}\,n_a~,\nonumber
\end{eqnarray}
\begin{eqnarray}
  \Lambda &=&
  (\Lambda\Lambda_{\sigma\omega})^2-(\Lambda\Lambda_\sigma)(\Lambda\Lambda_\omega)~,
\end{eqnarray}
$n_a =\TGint_a  f_a$ is the number density of the  particle
species ``$a$'' and $\rho\scid_a =\TGint_a \frac{m^*_a}{E_a}\,
f_a$ is the scalar density. All integrals in matrix $\Lambda$ are
calculated in the rest reference frame of the fluid with the local
equilibrium distribution function. With the help of these
expressions the variation $\delta T^{ii}$ can be expressed as
\begin{eqnarray}
 \label{deltaP}
  \delta T^{ii}[\delta f_a]&=&3\sum_a\TGint_a F_a(\vec{p}^{\,2})\,\delta f_a~,
\end{eqnarray}
with
\begin{eqnarray}\label{Fa}
F_a(\vec{p}^{\,2})&=&\frac{\vec{p}^{\,2}}{3E_a}-K_\sigma
\frac{\pd\varepsilon_a}{\pd\sigma}-
  K_\omega\frac{\pd X^0_a}{\pd\omega_0}~,
\end{eqnarray}
\begin{eqnarray}
K_\sigma&=&\kappa\,\Lambda_\sigma -\frac{\pd
V}{\pd\omega_0}\,\Lambda_{\sigma\omega}~,\nonumber
\\
K_\omega&=&\kappa\,\Lambda_{\sigma\omega}-\frac{\pd
V}{\pd\omega_0}\,\Lambda_\omega~,\nonumber
\end{eqnarray}
\begin{eqnarray}
\kappa&=&\frac{\rmd  U}{\rmd\sigma}-\frac{\pd V}{\pd\sigma}
+\,\frac13\sum_a m_a^*\,\frac{\rmd
m_a^*}{\rmd\sigma}\TGint_a\frac{\vec{p}^{\,2}}{E_a^3}\,f_a^{\rm
l.eq.}~.\nonumber
\end{eqnarray}

Using Eqs. (\ref{vis}), (\ref{delf}), (\ref{deltaP}) we
obtain
\begin{eqnarray}\label{bulkSHMCour}
  \zeta &=& -\frac{1}{T}\sum_a\avgint{\tau_a
  Q_a(\vec{p}^{\,2})F_a (\vec{p}^{\,2})}~.
\end{eqnarray}
Setting $F_a =\vec{p}^{\,2}/3E_a$, see (\ref{Fa}), we reproduce the
result (\ref{viskold}).

If the Landau-Lifshitz conditions (\ref{LLaI2}), (\ref{LLaI1})
are not fulfilled with the particular
distribution (\ref{delf}), we may still fulfill them
doing
the shift
\begin{eqnarray}
\label{shiftaI}
 \tau_a Q_a(\vec{p}^{\,2})\rightarrow \tau_a
 Q_a(\vec{p}^{\,2})+y^B\,t_a^B +y^St_a^S +x\,\varepsilon_a,
\end{eqnarray}
where $x$ and $y^{B,S}$ are some constants. These constants are
associated with the conservation of the energy and  the baryon and
strange charges. Values of $y^B$ and $y^S$ are similar to  baryon and
strange chemical potentials.  Since $\mu_S\neq 0$ even for hadron
matter with zero net strangeness, we cannot exclude the term
$y^S$.

If one considers only elastic scattering  of particles described
by the exact Boltzmann collision term,  the replacement
(\ref{shiftaI}) is fully legitimate since it generates new
solutions of the original Boltzmann equation, see \cite{CK10}.
However, one can show that for multi-particle systems considered
within the relaxation time approximation even with
energy-averaged values of $\tau_a$, the above replacement does not
result in new solutions. Even for one species but with the
energy-dependent relaxation time $\tau (E^{\rm l.eq.})$, the  replacement
does not generate new solutions. Thus, {\em we actually fulfill
the Landau-Lifshitz conditions at the price that the solutions of
the Boltzmann equations with the collision terms~(\ref{trel})
might be spoiled.}

 After performing the replacement
(\ref{shiftaI}) in the conditions (\ref{LLaI2}), (\ref{LLaI1}), we
arrive at the system of linear equations for $x$ and $y$. Finally,
we obtain
\begin{eqnarray}
\label{zetaaI}
    \zeta&=&-\frac1T\sum_a\left<\tau_a Q_a(\vec{p}^{\,2})\right.\nl
    &&\left.\times\left[F_a(\vec{p}^{\,2})-\gamma\varepsilon_a-t_a^B\chi_B -t_a^S\chi_S\right]\right>,
\end{eqnarray}
where
\begin{eqnarray}
\gamma J&=&\begin{vmatrix}
  \sum_a\avgint{\varepsilon_a
F_a(p)} & a_{12} & a_{13} \\
  \sum_at_a^B\avgint{F_a(p)} & a_{22} & a_{23} \\
  \sum_at_a^S\avgint{F_a(p)} & a_{23} & a_{33} \\
\end{vmatrix},\nl
\chi_B J&=&\begin{vmatrix}
  a_{11}&\sum_a\avgint{\varepsilon_a
F_a(p)} & a_{13}\\
  a_{12}&\sum_at_a^B\avgint{F_a(p)} & a_{23}   \\
  a_{13}&\sum_at_a^S\avgint{F_a(p)} & a_{33}  \nonumber\\
\end{vmatrix},\\
\chi_SJ&=&\begin{vmatrix}
  a_{11}&a_{12}&\sum_a\avgint{\varepsilon_a
F_a(p)} \\
  a_{12}&a_{22}&\sum_at_a^B\avgint{F_a(p)}   \\
  a_{13}&a_{23}&\sum_at_a^S\avgint{F_a(p)} \\
\end{vmatrix},
\end{eqnarray}
\begin{eqnarray}
(a_{ij})\!\!&=&\!\!\!
\begin{pmatrix}
  \sum_a\avgint{\varepsilon_a^2} & \sum_at_a^B\avgint{\varepsilon_a} & \sum_at^S_a\avgint{\varepsilon_a} \\
  \sum_at_a^B\avgint{\varepsilon_a} & \sum_a(t_a^B)^2\avgint{1} & \sum_at_a^B t_a^S\avgint{1} \\
  \sum_at^S_a\avgint{\varepsilon_a} & \sum_at_a^Bt_a^S\avgint{1} & \sum_a(t^S_a)^2\avgint{ 1} \nonumber\\
\end{pmatrix}\!,
\end{eqnarray}
\begin{eqnarray}
J&=&\det \|a_{ij}\|.\nonumber
\end{eqnarray}
We stress that, as in model I, all quantities in Eqs.
(\ref{bulkSHMCour}) and (\ref{zetaaI}) including $E_a$ are taken
at local equilibrium.

\subsection{ Model III (Ref. \cite{CK10}).} Above we assumed that
the original Boltzmann collision terms ${\rm St}f_a^{\rm
l.eq.}(E_a^{\rm l.eq.})=0$. However, at calculating the viscosity
coefficients in the quasiparticle Fermi liquid theory, one often
uses \cite{AKh} that similar equality is valid also for $E_a$,
${\rm St}f_a^{\rm l.eq.}(E_a)=0$, being a functional of exact
non-equilibrium distribution functions. This is so because the
energy conservation $\delta$-function pre-factors depend on exact
particle energies. Thus, {\em{ in the relaxation time
approximation one can write}}
 \begin{eqnarray}\label{trelChK}
 {\rm{St}}
f_a =-\delta \widetilde{f}_a/\widetilde{\tau}_a  ,\quad \delta
\widetilde{f}_a =f_a -f_a^{\rm l.eq.}(E_a),
\end{eqnarray}
and thereby
\begin{align} \delta
f_a
= \delta\tilde
f_a+\frac{\pd f_a}{\pd\varepsilon_a}\,\left[\frac{\pd
\varepsilon_a}{\pd\sigma}\,\delta\sigma+\frac{\pd
X^0_a}{\pd\omega_0}\,\delta\omega_0\right].
\end{align}
 The relaxation time $\widetilde{\tau}_a$ is
 in general different from $\tau_a$ introduced above.
Note that due to the smallness of $\delta \tilde f_a$, one can
consider here $\widetilde{\tau}_a=\widetilde{\tau}_a(E_a^{\rm
l.eq.})$ as a function of the energy $E_a^{\rm l.eq.}$.

 The difference between the approach \cite{CK10}
 and those exploited in  Section 3.2
 is that here following
 \cite{CK10} we express all variations  through $\delta\tilde  f_a $ and
 $\widetilde{\tau}_a $,
 which now depend on
 $f_a^{\rm l.eq.} (E_a)$,
 rather than through $\delta f_a $ and $\tau_a$ depending on $E_a^{\rm l.eq.}$.
  Since $E_a$ are fixed and thus not varied, we get the same
expression for the value $\delta T^{ii}$ as (\ref{deltaPaIII}) but
 with the substitution $\delta\tilde f_a$ instead of $\delta
f_a$.
 Here one should note that in Section 3.1 
the quantity $E_a [f_a]$  was not varied according to our additional assumption, while now the expression for $\delta
T^{ii}[\delta \widetilde{f}_a]$ becomes a fully correct relation.

Within the relaxation time approximation using~(\ref{trelChK}) we
arrive at 
\begin{eqnarray}\label{viskoldCK}
  \zeta &=& -\frac1T\sum_a\avgint{\widetilde{\tau}_a
  Q_a(\vec{p}^{\,2})\frac{\vec{p}^{\,2}}{3E_a}}~,
\end{eqnarray}
i.e., we arrived at   expression (\ref{viskold}), where
$\tau_a$ is replaced by $\widetilde{\tau}_a$ and $E_a^{\rm l.eq.}$
is replaced by exact value $E_a$.
The later difference in $\zeta$ is not essential since  $E_a$ and $E_a^{\rm
l.eq.}$ differ only in terms linear in the frame velocity
gradients. Note that in practical calculations the relaxation time
is evaluated with the help of phenomenological particle cross
sections. In this case one cannot distinguish $\tau_a$ and
$\widetilde{\tau}_a$.

  Eqs. (\ref{viskold}), (\ref{bulkSHMCour}),
(\ref{viskoldCK}) demonstrate differences between values of the
bulk viscosities calculated in our three models before the
Landau-Lifshitz matching conditions are used. Now let us exploit
Landau-Lifshitz  conditions in our model III. Performing
similar calculations to those we have done in the previous section,
we rewrite the Landau-Lifshitz conditions as
\begin{eqnarray}
 \label{LLKap_2}\sum_a\avgint{\widetilde{\tau}_aQ_a\left(t_a^{B,S}-\frac{\pd
\varepsilon_a}{\pd\mu_{B,S}}\right)}&=&0 ~,\\
\sum_a\left<\widetilde{\tau}_a Q_a\left(\varepsilon_a-T\frac{\pd
\varepsilon_a}{\pd T}-\mu_B\frac{\pd \varepsilon_a}{\pd\mu_B}
\right.\right.\nl\left.\left. -\mu_S\frac{\pd
\varepsilon_a}{\pd\mu_S}\right)\right>&=&0~.\label{LLKap_1}
\end{eqnarray}
The integration is performed at fixed $E_a$ rather than
$E_a^{\rm l.eq.}$.  Making use of the shift
\begin{eqnarray}
 \widetilde{\tau}_a Q_a(\vec{p})\rightarrow \widetilde{\tau}_aQ_a(\vec{p})+y^B\,t_a^B+y^S\,t_a^S +x\,\varepsilon_a
\end{eqnarray}
and solving the corresponding system of linear equations for $x$,
$y^B$ and $y^S$ we find
\begin{equation}
\label{zetaKap}
    \zeta_{\rm ChK} =\frac1T\sum_a\avgint{ \widetilde{\tau}_a Q^2_a(\vec{p})}~.
\end{equation}
Eq. (\ref{zetaKap}) formally coincides with that obtained in
\cite{CK10} but
new terms depending on $\mu_{B,S}$ are involved, cf.
(\ref{defQa}). We note that, as in  Section 3.1, 
{\em{ we actually fulfill the Landau-Lifshitz conditions but spoil the solutions of the Boltzmann equations with
the collision terms (\ref{trelChK}).}}

Note that for  hydrodynamical calculations one needs
values of viscosities computed at local equilibrium, whereas in
the model III all quantities including resulting value of the
viscosity implicitly depend on non-equilibrium values $E_a$ rather
than on $E_a^{\rm l.eq.}$. However corrections to the viscosities
due to difference between $E$ and $E_a^{\rm l.eq.}$ are linear in
the velocity gradients and can be neglected.

It is worthwhile  to  emphasize some properties of derived
expressions (\ref{viskold}), (\ref{bulkSHMCour}),
(\ref{viskoldCK}) (when fulfillment of the Landau-Lifshitz
conditions is not yet implied), Eq. (\ref{bulkSHMC}) (if they
are implied but not checked), and Eqs. (\ref{zetaaI}) and
(\ref{zetaKap}) (when Landau-Lifshitz conditions are fulfilled).
The expression (\ref{zetaKap}) has an additional advantage that it
is explicitly positively definite. Positive definiteness of the
expression (\ref{zetaaI}) can be proven  at least for
one-component matter with a temperature-dependent quasiparticle
mass and for zero chemical potential. Energy dependence of $\tau $
does not spoil the proof. Positive definiteness of
(\ref{bulkSHMC}) can be proven if additionally one assumes $\tau
=\,$const. These proofs have been performed in \cite{GBulk1}.
Also, we have checked numerically that quantities
(\ref{bulkSHMC}) and (\ref{zetaaI}) are positive  for our model in
the whole region of the parameters used in contrast with quantities
(\ref{viskold}), (\ref{bulkSHMCour}), (\ref{viskoldCK}) which are
positive in the limited region of parameters. Definitely, these
values can be used for estimations only in regions, where they are
positive.

 In
the ideal gas (IG) limit, i.e., if we put $\ X_a^0=0$ and assume
$m_a$ to be temperature independent,  Eqs. (\ref{zetaaI}) and
(\ref{zetaKap}) exactly reproduce the standard expression, which
is the sum of quadratic terms. Also, expressions
(\ref{viskold}), (\ref{bulkSHMCour}), (\ref{viskoldCK}) and  Eq.
(\ref{bulkSHMC}) reproduce the IG limit, if it is additionally
assumed $\tau_a=\tau=\,$const (e.g., for one-component system with
$\tau (E) =\,$const, cf. \cite{G85} in case of the pion
gas).

\section{Numerical results} Details of calculations of the
relaxation time $\tau_a$ and of  shear and bulk viscosities in our
SHMC model  can be found in \cite{SHMCbulk}. Since in such a
complicated multi-particle system, as we study, values $\tau_a$
cannot be calculated microscopically  but  can  only be evaluated
using empirical values of the cross sections, we cannot
distinguish $\tau_a$ and $\widetilde{\tau}_a$ and therefore, as in
\cite{GBulk1}, we consider them to be the same.
\begin{figure}[t]
\centerline{\includegraphics[width=0.7\textwidth,clip]{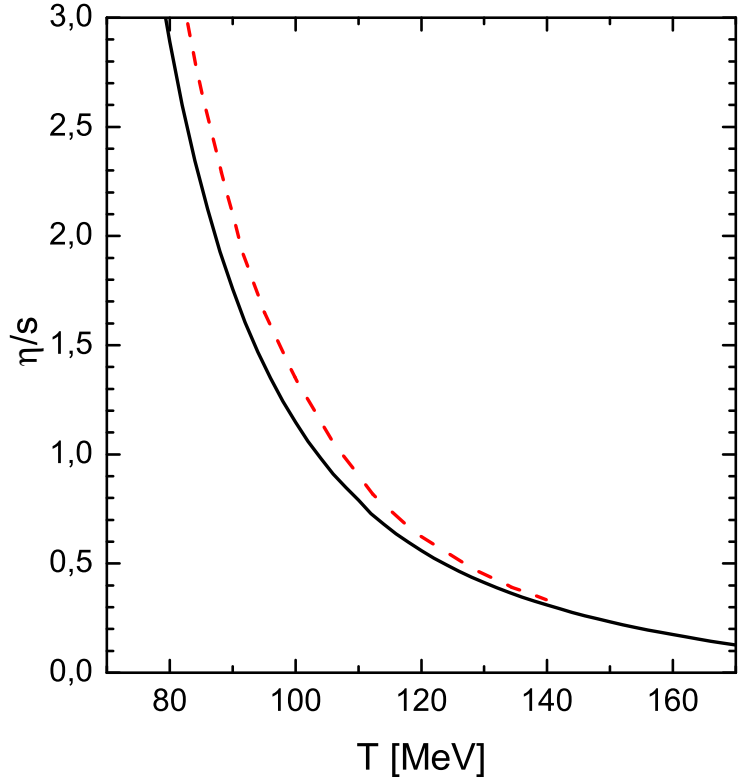}}
\caption{(Color online.) The shear viscosity to
entropy density ratio as a function of the temperature for $\mu_B =
0$. The solid line is the result of the given work and the
long-dashed line is that   for the interacting pion gas
\cite{Weise}.
 }
 \label{eta0}
\end{figure}

In Fig. \ref{eta0} we  compare the ratio of the shear viscosity to
the entropy density  at $\mu_B=0$, being calculated in the SHMC model
of \cite{SHMCbulk} used in the present work, and that computed  in
recent paper \cite{Weise} in the model of the interacting pion
gas. The results demonstrate rather appropriate overall agreement
although in the SHMC-model the relaxation time is evaluated using
phenomenological values of the cross sections of particle species
and the calculation of the viscosity is performed in the
relaxation time approximation, whereas in \cite{Weise} the
cross sections of the processes are computed (but only for pions)
and the viscosity of the pion  gas is estimated with the help of
the Kubo formalism. This agreement might be considered as an
additional argument in favor of estimates of the relaxation time
used in \cite{SHMCbulk} and in the given work for $\mu_B=0$ and
$T\lsim 150$~MeV, when  pions are dominating species.

\begin{figure}[t]
\centerline{\includegraphics[width=0.7\textwidth,clip]{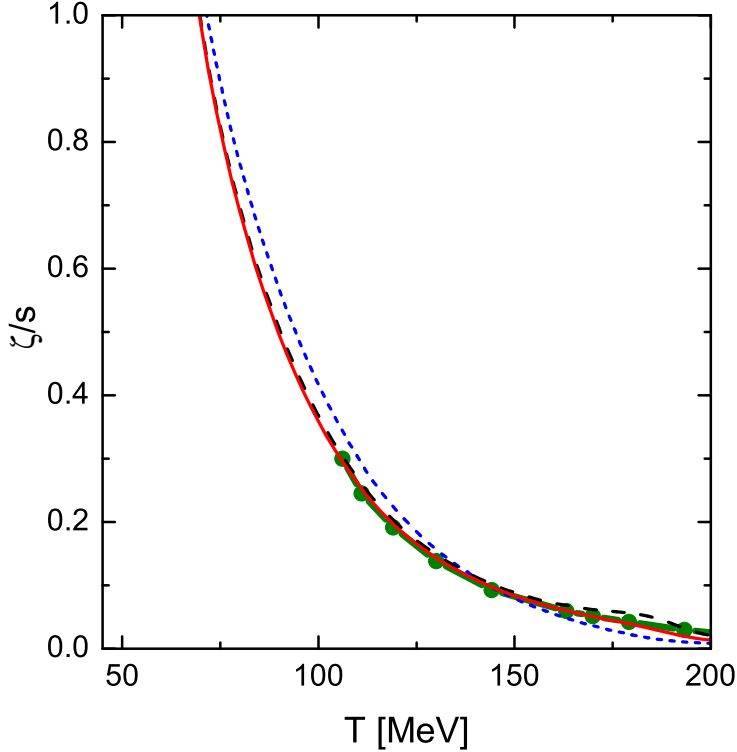}}
\caption{(Color online.) The ratio of the bulk viscosity to the
entropy density  as a function of temperature at $\mu_B=0$. The
solid line is the result of the given work, where $\zeta$ satisfies Eq.
(\ref{zetaaI}); the short-dashed line is taken from
\cite{SHMCbulk}, where $\zeta$ fulfills Eq.~(\ref{bulkSHMC}); and the
long-dashed line is the result of \cite{CK10}, with $\zeta$ satisfied
Eq.~(\ref{zetaKap}). The circles present the linear
sigma model \cite{Dobado}.
 }
 \label{zetas0}
\end{figure}

In Fig. \ref{zetas0}, we show the bulk viscosity to the entropy
density ratio at $\mu_B=0$ as a function of the temperature. The
solid line is calculated in the given work
following Eq. (\ref{zetaaI}), model II. The long-dashed
curve is the result of Eq. (\ref{zetaKap})  (our model III)
derived with the ansatz \cite{CK10}. In order to
perform this calculation we replaced the exact $E$ with the local
equilibrium value $E^{\rm l.eq.}$ in (\ref{zetaKap}). The
short-dashed curve is our old result \cite{SHMCbulk} calculated
following Eq. (\ref{bulkSHMC}), model I. We see that all
three results (especially those calculated following Eqs.
(\ref{zetaaI}) and (\ref{zetaKap})) are close to each other for
temperatures $T\lsim 150$~MeV. For higher temperatures, deviations
become a little bit more pronounced. Comparing our results with
those computed recently in the  linear sigma model \cite{Dobado}
in the crossover region (line with circles in the figure) one may
see an overall agreement, although models as well as  final
expressions used for the bulk viscosities are essentially
different.
\begin{figure}[t]
\includegraphics[width=\textwidth,clip]{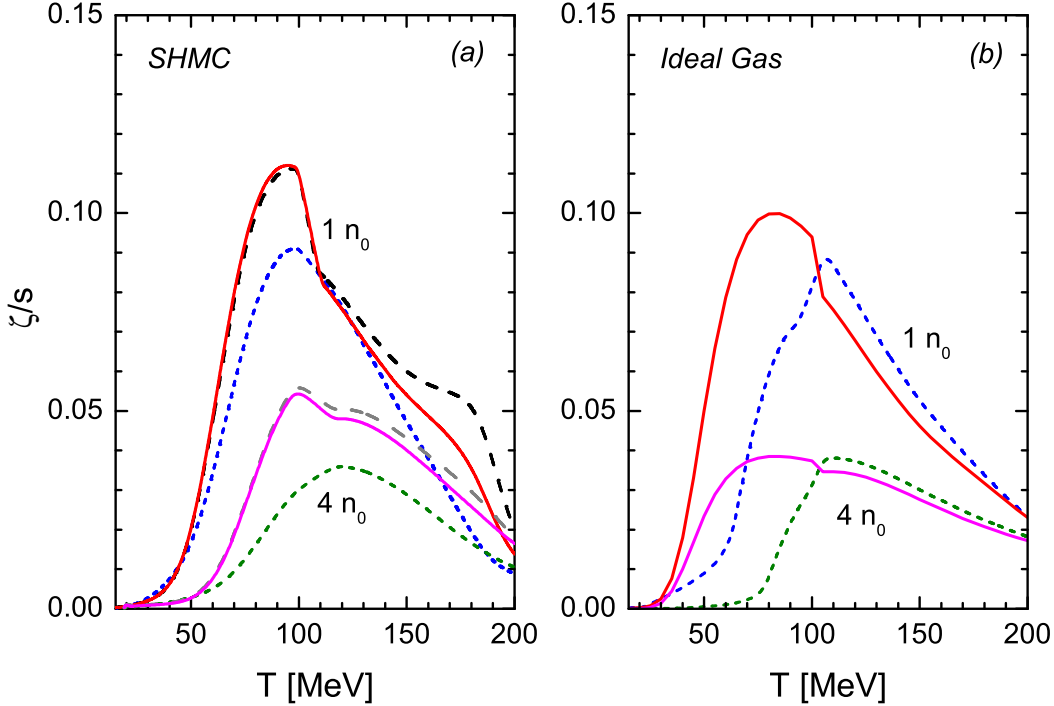}
\caption{(Color online.) The ratio of the bulk viscosity to the
entropy density as a function of temperature for $\mu_B\neq 0$,
at  two values of the baryon density $n_B/n_0=1$ and $4$ (from
top to bottom). Notation is the same as in Fig. \ref{zetas0}.
 }
 \label{zetaeta0}
\end{figure}

In Fig. \ref{zetaeta0}, the ratio of the bulk viscosity to the
entropy density is presented as a function of temperature for two
values of the baryon density $n_B=n_0$ and $4n_0$, where
$n_0=0.16$~fm$^{-3}$ is the nuclear density at the saturation
point. The notation is the same as in  Fig. \ref{zetas0}. The
results are shown for the SHMC model (a) and for the IG model (b)
with the same hadron set. We see that the curves calculated by Eq.
(\ref{zetaaI}) and Eq. (\ref{zetaKap}), where we again
replaced $E$ with $E^{\rm l.eq.}$, are closer to each other than to
the curve calculated with Eq. (\ref{bulkSHMC}). For the IG, the
(solid and long-dashed) curves calculated following Eqs.
(\ref{zetaaI}) and (\ref{zetaKap}) coincide.  Also, these curves
are  close to those estimated according to Eq.~(\ref{bulkSHMC})
for $T\gsim 100$~MeV. Comparing figures (a) and (b) we conclude
that the presence of the quasiparticle interaction is more
significant for low temperatures ($T \lsim 100$~MeV) and it
becomes less important for higher temperatures.

As we see from Figs. \ref{zetas0} and \ref{zetaeta0}, within our
SHMC model  all three results (\ref{zetaaI}), (\ref{zetaKap}), and
~(\ref{bulkSHMC}) yield positive values in the temperature-density
region  of interest. Also we checked positive definiteness of these expressions in
case of  the ordinary RMF Walecka model by switching off the hadron mass-coupling
scaling.

{\bf{Conclusion.}}
 Due to the complexity of the hadron system formed in actual
 heavy-ion collisions, it is hard to calculate the viscosity
coefficients from the first principles with
the help of  the Kubo formulas.  One could use the  Kadanoff-Baym
kinetic equations for the hadron resonances to derive general
expressions for the kinetic coefficients,  but at present
realistic calculations do  not seem  possible even in the
relaxation time approximation \cite{V11}. Therefore, making use of
the quasiparticle Boltzmann-like equations (being treated within
the relaxation time approximation) can be considered as a forced
step for practical evaluations of the kinetic coefficients in the
given problem. The scaling hadron mass-couplings model of Ref.
\cite{SHMCmodel} is an appropriate tool for the description of
the  equation of state of the hot and dense hadron system. The
knowledge of the latter is necessary  in order  to perform
evaluations of the kinetic coefficients.
 However, even an application of simplified phenomenological
expressions for the relaxation times of different species does not
allow one to proceed in calculation of the bulk viscosity without
doing additional assumptions, in particular the Landau-Lifshitz
conditions should be fulfilled. However, these conditions cannot be
satisfied on the class of solutions of the Boltzmann equations for
our multi-component system treated within the relaxation time
approximation, and additional ansatze are needed. Contrary, the
result for the shear viscosity proves to be rather robust to these
reductions, see \cite{GBulk1}.

Thus, we studied three models  (models I and III have been
previously used in the literature) and derived three expressions
for the bulk viscosity  (\ref{bulkSHMC}), (\ref{zetaaI}), and
(\ref{zetaKap}), generalized to the case of nonzero chemical
potentials $\mu_{B,\,S}$. In derivation of (\ref{bulkSHMC})
fulfillment of the Landau-Lifshitz conditions is just implied,
whereas (\ref{zetaaI}), and (\ref{zetaKap}) fulfill these
conditions. Luckily, numerical evaluations, shown in  Figs.
\ref{zetas0} and \ref{zetaeta0}, carried out following all three
expressions deviate not much from each other and confirm earlier
results~\cite{SHMCbulk,CK10,GBulk,GBulk1,BKR10}. Although these
results can be considered only as rough estimations, Eq. (\ref{zetaaI}) and Eq. (\ref{zetaKap}) seem to be
more theoretically justified, while Eq. (\ref{bulkSHMC}) is less
established. In order to perform more accurate calculations, one
should go beyond the scope of the relaxation time approximation
and fulfill the Landau-Lifshitz conditions on the class of
solutions of the kinetic equation. However, such calculations are
much more involved than the estimations presented in the given
work and have not yet been carried out for multi-component systems
with strong interactions.


{\bf Acknowledgements.} This work was supported by RFBR Grant No.
11-02-01538-a.

\end{document}